\documentclass[twocolumn,twoside,5p,preprint]{elsarticle}
\biboptions{sort&compress}

\usepackage[T1]{fontenc}
\usepackage[utf8]{inputenc}
\usepackage[english]{babel}
\usepackage{amssymb,amsthm,amsmath,amstext,amsbsy,amsopn}
\usepackage{bbm}
\usepackage{graphicx}
\usepackage{isotope}
\usepackage{float}
\usepackage[dvipsnames]{xcolor}
\usepackage{orcidlink}
\usepackage{hyperref}
\hypersetup{
    colorlinks = true,
    citecolor  = blue,
    linkcolor  = blue,
    urlcolor  = blue
}

\newcommand{\la}{\langle}
\newcommand{\ra}{\rangle}
\newcommand{\Xmax}[1]{\ensuremath{#1_\text{max}}}

\newcommand{\cluster}[1]{\ensuremath{\mathcal{T}_{#1}}}

\newcommand{\elem}[2]{{$^{#2}$}\text{#1}}

\newcommand{\ai}{\textit{ab initio}}
\newcommand{\ie}{\textit{i.e.}}
\newcommand{\eg}{\textit{e.g.}}

\newcommand{\MeV}{\text{MeV}}
\newcommand{\fm}{\text{fm}}

\newcommand{\deltago}{$\Delta$NNLO$_\text{GO}$}
\newcommand{\magicint}{1.8/2.0~(EM)}
\newcommand{\emarthuis}{1.8/2.0~(EM7.5)}

\begin{document}

\allowdisplaybreaks

\title{
\textit{Ab initio} calculations of nuclear charge radii across and beyond $^{132}$Sn: \\ Putting chiral EFT nuclear interactions to the test}
\address[kul]{KU Leuven, Instituut voor Kern- en Stralingsfysica, 3001 Leuven, Belgium}
\address[ulb]{Institut d’Astronomie et d’Astrophysique, Université Libre de Bruxelles, 1050 Brussels, Belgium}
\address[blu]{Brussels Laboratory of the Universe – BLU-ULB, 1050 Brussels, Belgium}
\address[tud]{Technische Universit\"at Darmstadt, Department of Physics, 64289 Darmstadt, Germany}
\address[emmi]{ExtreMe Matter Institute EMMI, GSI Helmholtzzentrum f\"ur Schwerionenforschung GmbH, 64291 Darmstadt, Germany}
\address[cea]{IRFU, CEA, Universit\'e Paris-Saclay, 91191 Gif-sur-Yvette, France}
\address[mpik]{Max-Planck-Institut f\"ur Kernphysik, Saupfercheckweg 1, 69117 Heidelberg, Germany}

\author[kul,ulb,blu]{P.~Demol \orcidlink{0000-0003-2511-7179}}
\ead{pepijn.demol@ulb.be} 

\author[tud,emmi,kul]{U.~Vernik \orcidlink{0000-0002-3717-945X}}
\ead{urban.vernik@tu-darmstadt.de}

\author[cea,kul]{T.~Duguet \orcidlink{0000-0002-7596-3851}}
\ead{thomas.duguet@cea.fr} 

\author[tud,emmi,mpik]{A.~Tichai \orcidlink{0000-0002-0618-0685}}
\ead{alexander.tichai@tu-darmstadt.de}

\begin{abstract}
Charge radii are investigated along the Tin isotopic chain via \ai{} Bogoliubov coupled cluster calculations at the singles and doubles level. In addition to the reproduction of absolute radii, the parabolic behavior of isotopic shifts between the $N=50$ and $N=82$ magic numbers and the kink through \elem{Sn}{132} are shown to provide stringent tests for state-of-the-art chiral effective field theory ($\chi$EFT) inter-nucleon interactions. Indeed, none of the employed fine-tuned interactions can capture all such key characteristics. Eventually, the pronounced sensitivity of the results to the employed Hamiltonian beyond \elem{Sn}{132} provides a unique playground to pin down critical attributes of $\chi$EFT inter-nucleon interactions in the future. This calls for measuring isotopic shifts both towards \elem{Sn}{100} and beyond \elem{Sn}{134}, as well as for performing high-accuracy \ai{} calculations of mean-square radii in heavy open-shell nuclei by adding both triples corrections to the many-body wave function and the two-body charge density correction to the operator.
\end{abstract}

\maketitle

\section*{Introduction}

The description of atomic nuclei rooted into the underlying theory of the strong force currently allows the \ai{} description of various nuclear observables in light and medium-mass systems, \eg{}, binding energies~\cite{Stroberg2021,Tichai18BMBPT,Hu2025texas}, charge radii~\cite{Novario2020a,Koszorus2021,Heinz2025calcium,Gustafsson2025,Wolfgruber2025,Mueller2025}, low-lying spectroscopy~\cite{Tichai2024dmrg,Heinz2025calcium}, collective excitations~\cite{Frosini2021mrIII,Hagen2022PCC}, electromagnetic responses and giant resonances~\cite{Bacca:2013dma,Bacca:2014rta,Simonis:2019spj,Acharya:2024xah,Porro:2024vlc,Porro:2024tzt,Porro:2024pdn,Porro:2024bid,Porro:2025lhp}, electro-magnetic moments and transitions~\cite{Mior16dipole,Miyagi2023currents,Bonaiti2024,Marino2025openshell}, weak decays~\cite{Gysb19beta} or the search for new physics~\cite{Heinz2025overlap,Door2025ytterbium}. Pushing forward the mass frontier~\cite{Morr17Tin,Hu2021lead,Arthuis2020a,Tichai2024bcc,Door2025ytterbium,Hebe203NF,Hebeler2023,Miyagi2021,Miyagi2026r4,Arthuis24a}, such first-principle calculations could recently address the doubly magic character of  $^{266}$Pb~\cite{Bonaiti:2025bsb} and compute the binding energy of all even-mass Sn isotopes~\cite{Tichai2024bcc}.

Based on mildly scaling many-body expansion methods~\cite{Herg20review}, inter-nucleon interactions based on chiral effective field theory ($\chi$EFT)~\cite{Epel2020prec,Machleidt2024} can thus be put to the test in a new mass regime, further deepening some of the challenges identified in lighter systems. A particularly daunting difficulty concerns the reproduction of nuclear charge radii, such as the infamous inverted parabolic behavior between \elem{Ca}{40} and \elem{Ca}{48} or the marked rise between the latter isotope and \elem{Ca}{52}~\cite{Ruiz16Calcium}. As a matter of fact, the challenge is even more severe given that commonly used interactions tend to systematically and significantly underpredict absolute charge radii~\cite{Lapoux:2016exf,Simo17SatFinNuc}; only recently have new interactions emerged that reproduce both binding energies and charge radii simultaneously in doubly magic nuclei over a significant mass range~\cite{Ekstrom:2015rta,Jian20N2LOGO,Arthuis24a,Hu2025texas}. 

In this context, the recent charge radius measurement of \elem{Sn}{134} signaling a pronounced kink through \elem{Sn}{132}~\cite{Gorges2019} and similar measurements in neutron-deficient isotopes~\cite{Gustafsson2025} make Sn isotopes a remarkable playground. The corresponding goal is to characterize the capacity of existing $\chi$EFT nuclear interactions to reproduce, and eventually predict, the evolution of nuclear charge radii over a long chain of heavy isotopes.

In this Letter, first-principle calculations of charge radii are performed for the even-mass \elem{Sn}{96-150} isotopes, going through and beyond $N=50$ and $N=82$ neutron magic numbers. Our detailed analysis reveals that accurately reproducing charge radii throughout the Sn isotopic chain constitutes a major challenge for state-of-the-art $\chi$EFT-based nuclear interactions.

\section*{Theoretical framework}

The many-body time-independent Schr\"odinger equation
\begin{align}
    H | \Psi_n \ra = E_n | \Psi_n \ra \, ,
\end{align}
is solved based on the nuclear Hamiltonian $H=T+V+W$, with $T$ the (intrinsic) kinetic energy, $V$ the two-body potential and $W$ the three-body potential.

The fully interacting nuclear ground state is parameterized in Bogoliubov coupled cluster (BCC) theory~\cite{Sign14BogCC,Henderson2014,Tichai2024bcc} as
\begin{align}
    |\Psi_0 \ra \equiv e^\cluster{} | \Phi \rangle \, ,
\end{align}
where  the many-body reference state $| \Phi \ra$ is chosen as a particle-number-breaking Bogoliubov vacuum obtained as the variational solution of the Hartree-Fock-Bogoliubov (HFB) mean-field equations~\cite{RingSchuck}, whereas \cluster{} denotes the (quasi-particle) cluster operator. The Hartree-Fock-Bogoliubov equations  are solved by enforcing rotational invariance, thus yielding a $J=0$ angular-momentum eigenstate  $| \Phi \ra$. The average particle (\ie{} proton or neutron) number is constrained to match the physical value by introducing a chemical potential $\lambda$ and the grand potential $\Omega \equiv H - \lambda A$, where $A$ is the particle (\ie{} proton or neutron) number operator. The grand potential $\Omega$ thus defines the fundamental input to the BCC many-body expansion~\cite{Soma13GGF2N,Sign14BogCC,Duguet2015u1,Tichai18BMBPT,Tichai2020review,Tichai2024bcc}.

Due to the nature of $| \Phi \ra$, the reference energy $E = \la \Phi | H | \Phi \ra$ already incorporates pairing correlations associated with the superfluid character of open-shell nuclei. Missing dynamical correlations $\Delta E_\text{BCC}$ are added via the action of the cluster operator $\mathcal{T}\equiv \mathcal{T}_{1}+\mathcal{T}_{2}+\mathcal{T}_{3}+\ldots$, where $\mathcal{T}_k$ is a $2k$-quasi-particle excitation operator. The associated unknown amplitudes are solved for iteratively by minimizing the BCC residual $\mathcal{R} \equiv  Q \tilde \Omega | \Phi \ra\la \Phi |$, where $Q$ projects on the manifold of single, double, \ldots quasi-particle excitations and $\tilde \Omega \equiv (e^{- \cluster{}} \,  \Omega \, e^\cluster{})_c$ denotes the similarity-transformed grand potential. 

The evaluation of a nuclear observable $O$ requires an explicit parametrization of the left ground state. This is achieved within a linear-response framework where the bra state is written in terms of a linear expansion
\begin{align}
    \la O \ra \equiv \la \Phi | (1 + \Lambda) \, e^{-\cluster{}} \,  O \, e^\cluster{} | \Phi \ra \, ,
\end{align}
where $\Lambda = \sum_n \Lambda_n$ denotes a quasi-particle de-excitation operator with components
\begin{align}
    \Lambda_{n} = \frac{1}{(2n)!} \sum_{p_1 ... p_{2n}} \Lambda_{p_1 ... p_{2n}} \,
    \beta_{p_{2n}} \cdots \beta_{p_1} \, .
\end{align}
While in principle the $\Lambda$ amplitudes have to be determined from a separate set of coupled equations~\cite{Shav09MBmethod}, they are presently approximated to first order in perturbation theory from the adjoint of the cluster operator, \ie{}, $\Lambda \equiv \cluster{}^\dagger$. For $O \equiv H$, the BCC correlation energy is obtained, \ie{}, $\la H \ra = \Delta E_\text{BCC}$.

In practice, both $\cluster{}$ and $\Lambda$ must be truncated to make the calculations numerically feasible. In this work, they are taken at the singles and doubles level, \ie{}, $\cluster{} = \cluster{1} + \cluster{2}$ and $\Lambda = \Lambda_1 + \Lambda_2$, giving rise to the BCCSD approximation. This is known to yield ground-state energies with a residual error due to missing triples $\cluster{3}$ of the order of $10 \%$ of $\Delta E_\text{BCC}$~\cite{Hage14RPP}. To reach a sub-percent accuracy on binding energies,  triples' correction must indeed be added~\cite{vernik26}. Since correlation effects are less important for nuclear radii, the BCCSD approximation is expected to be much more accurate for this observable. Furthermore, differential quantities such as two-neutron separation energies and isotope shifts largely benefit from error cancellations between correlated uncertainties in neighboring systems.

\begin{figure}[t!]
    \centering
    \includegraphics[width=0.95\columnwidth]{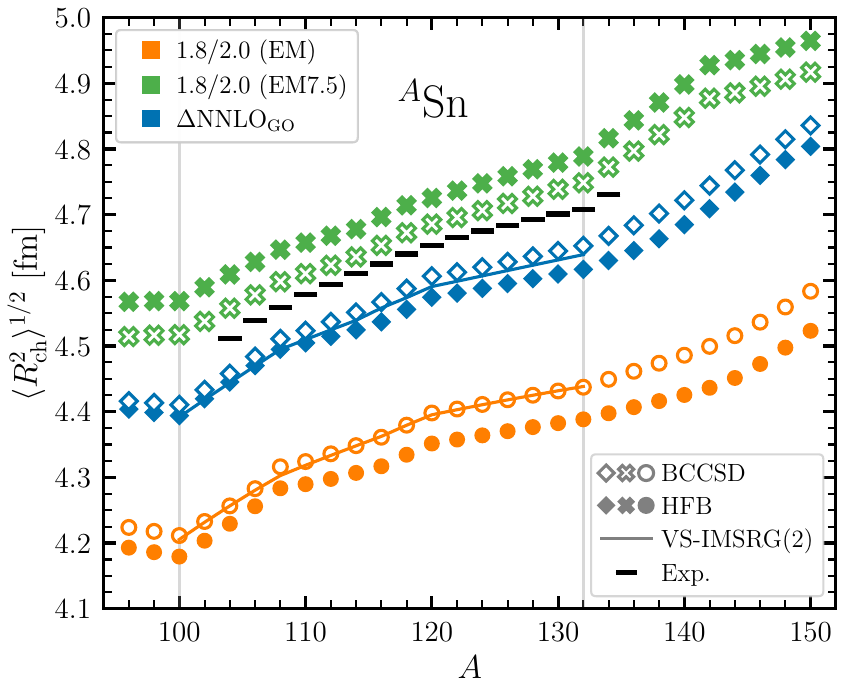}
    \caption{Theoretical and experimental charge radii. Hartree-Fock-Bogoliubov and BCCSD results in \elem{Sn}{96-150} are displayed for  the \magicint{}, \deltago{} and \emarthuis{} $\chi$EFT-based Hamiltonians. Valence-space IMSGR(2) results in  \elem{Sn}{100-132} are shown for the first two Hamiltonians~\cite{Miya25privcom,Gustafsson2025}.}
    \label{fig:obsabs}
\end{figure}

Three different fine-tuned $\chi$EFT-based Hamiltonians are employed in the present study
\begin{itemize}
\item The \magicint{} Hamiltonian~\cite{Hebe11fits} that yields binding energies with an accuracy better than $2\%$ over mid-mass nuclei~\cite{Stroberg2021} but underpredicts absolute charge radii by about $5\%$.
\item The $\Delta$-full \deltago{} Hamiltonian~\cite{Jian20N2LOGO} with a cutoff of $\Lambda = 394 \, \MeV$ that is adjusted on saturation properties of symmetric nuclear matter to better reproduce absolute charge radii.
\item The recently proposed \emarthuis{} Hamiltonian~\cite{Arthuis24a}, with a refitted value of the three-body low-energy constant $c_D = 7.5$ to match the ground-state energy and charge radius of \elem{O}{16}. This yields a reproduction of binding energies similar to the \magicint{} Hamiltonian while drastically improving absolute charge radii in doubly closed-shell nuclei.
\end{itemize}

\begin{figure*}[t!]
    \centering
    \includegraphics[height=7cm]{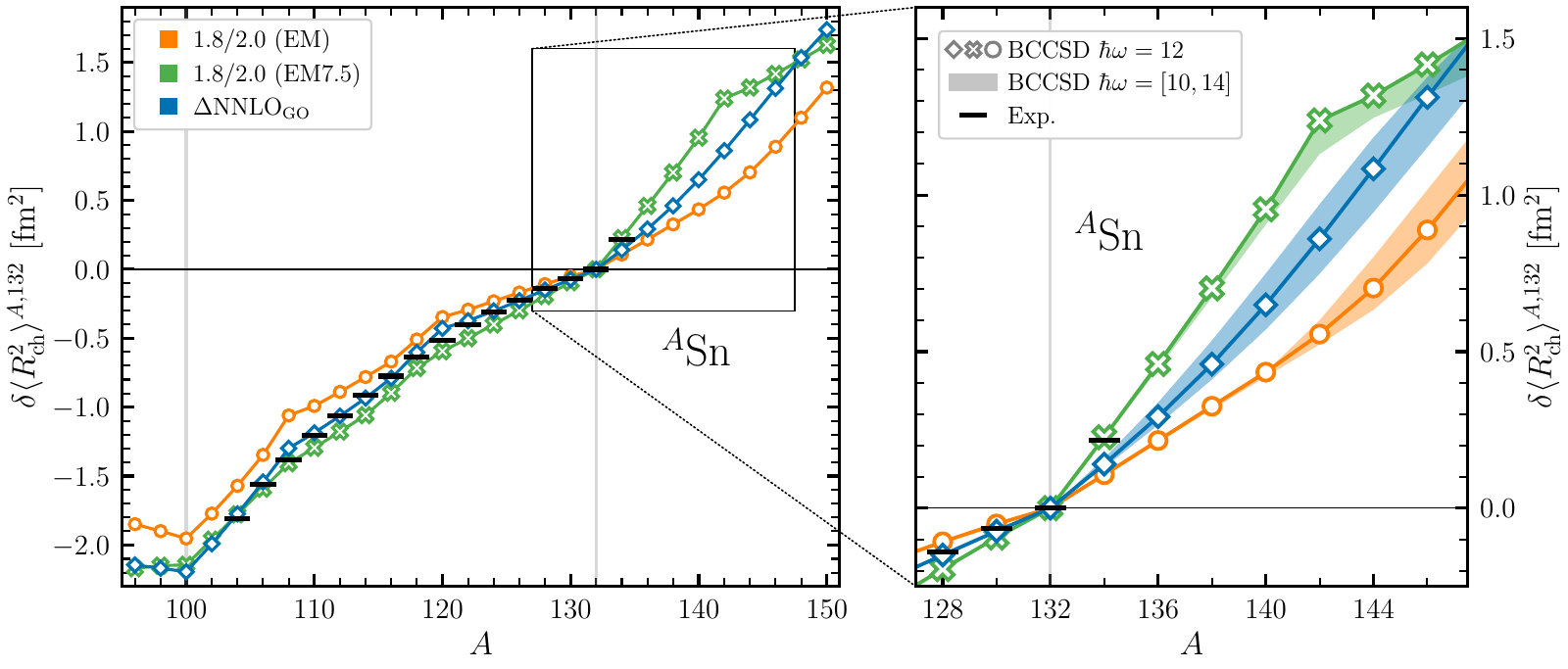}
    \caption{Theoretical and experimental isotopic shifts. The left panel shows BCCSD results in \elem{Sn}{96-150} for the \magicint{}, \deltago{} and \emarthuis{} $\chi$EFT-based Hamiltonians. In the right panel, a focused view of \elem{Sn}{128-146}  is presented, with colored bands displaying the HO frequency variation ($\hbar \omega = [10,14]$). }
    \label{fig:tin_isotopeshift}
    \label{fig:tin_isotopeshift_zoom}
\end{figure*}

Many-body operators are represented in a spherical harmonic oscillator (HO) basis including 13 major shells, \ie{}, $\Xmax{e} = (2n + l)_\text{max}=12$. The HO frequency $\hbar\omega=12$ presently employed is optimal for charge radii in this mass region~\cite{Arthuis24a,Art26privcom}. Model-space uncertainties are gauged by $i)$ varying $\hbar\omega$ over the interval $[10,14]$ (at $\Xmax{e} =12$) and by $ii)$ varying $\Xmax{e}$ from $12$ to $14$ (at $\hbar\omega=12$). Three-body matrix elements are further truncated at $E^{(3)}_\text{max} = e_1 + e_2 + e_3 = 24$, which is sufficient to reach converged results in the target mass regime~\cite{Miyagi2021,Tichai2024bcc}.
Nuclear matrix elements are obtained using the \textsc{NuHamil} code~\cite{Miyagi2024nuhamil}.

The mean-square charge radius is evaluated as
\begin{align}
    R^2_\text{ch} \equiv R_p^2 + \la r^2_\text{p} \ra + \la r^2_\text{n} \ra \cdot \frac{N}{Z} + \la r^2\ra_\text{so} + r_\text{DF} \, ,
\end{align}
where $R_p^2$ denotes the point-proton radius (including center-of-mass correction), $\la r_p^2 \ra = 0.771\, \fm^2$ the proton radius, $\la r_n^2\ra = -0.1149 \, \fm^2$ the neutron radius, $\la r^2\ra_\text{so}$ the spin-orbit correction~\cite{Ong10spinorbit,Heinz2025calcium} and $r_\text{DF} = 3/(4m_p^2c^4) = 0.033 \, \fm^2$ the relativistic Darwin-Foldy term~\cite{Fria97foldyShift}.

\begin{figure*}[t!]
    \centering
    \includegraphics[width=0.8\textwidth]{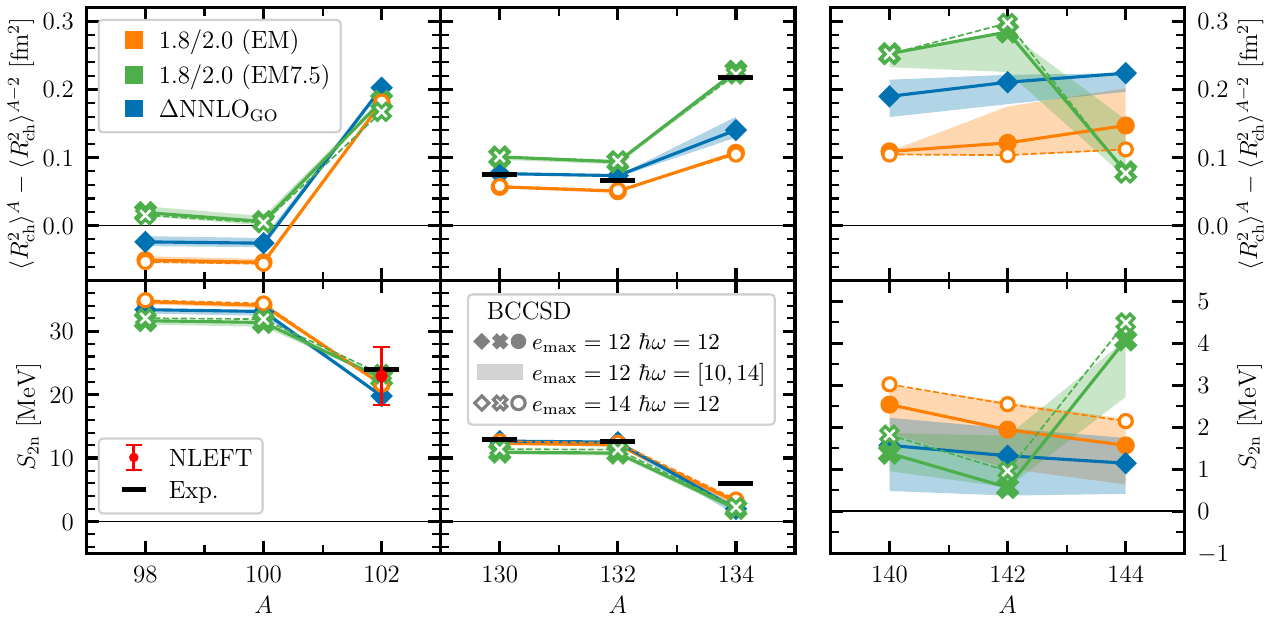}
    \caption{Differential isotopic shift (top) and two-neutron separation energies (bottom) across \elem{Sn}{100} (left), \elem{Sn}{132} (middle) and \elem{Sn}{142} (right). The two-neutron separation energy in \elem{Sn}{102} from  NLEFT calculations~\cite{hild25NLEFTSn} is also displayed in the bottom panel.}
    \label{fig:shellclosure}
\end{figure*}

Throughout the paper, our BCCSD results obtained with the \magicint{} and \deltago{} Hamiltonians are compared in the interval \elem{Sn}{100-132} to charge radii obtained from valence-space in-medium similarity renormalization group calculations truncated at the two-body level (VS-IMSRG(2)) and employing a \elem{Sn}{100} core~\cite{Miya25privcom,Gustafsson2025}. In Fig.~\ref{fig:shellclosure}, the two-neutron separation energy computed in \elem{Sn}{102} from nuclear lattice effective field theory (NLEFT)~\cite{hild25NLEFTSn} is also used as a benchmark.

\section*{Absolute charge radii and isotopic shifts}

Total nuclear charge radii are displayed in Fig.~\ref{fig:obsabs} between \elem{Sn}{96} and \elem{Sn}{150}. Experimental data are available over the large set of \elem{Sn}{104-134} isotopes. Noticeably, charge radii in  neutron deficient  \elem{Sn}{104-107} isotopes~\cite{Gustafsson2025} and in the neutron rich \elem{Sn}{134} isotope~\cite{Gorges2019} have been measured recently from high-resolution laser spectroscopy at ISOLDE.

First, BCCSD charge radii computed from the \magicint{} and \deltago{} Hamiltonians are seen to be highly consistent with VS-IMSRG(2) results, with a root-mean-square (rms) deviation of $0.1\%$ and $0.3\%$, respectively. Second, while radii obtained from the \magicint{} Hamiltonian significantly underestimate experimental data ($5.6\%$ rms error), improved saturation properties of the \deltago{} interaction and the fine-tuning of the \emarthuis{} Hamiltonian largely correct for this deficiency, reducing the rms error to $1.2\%$ and $0.7\%$, respectively. These results are consistent with those obtained earlier in doubly closed-shell nuclei~\cite{Arthuis24a}.

Although non negligible, BCCSD correlations contribute for only about $1\%$ of the total charge radius\footnote{While this is much less than for binding energies, it must be noticed that the next order contribution to charge radii from triples excitations ($0.7\%$) is only mildly suppressed in  \elem{Ca}{48}~\cite{Bon26privcom}. Consequently, the systematic convergence of the correlation expansion for charge radii deserves detailed investigations in the future.}. Interestingly, correlations always improve the agreement with experiment, \ie{} correlations systematically increase charge radii for the \magicint{} and \deltago{} Hamiltonians for which the mean-field baseline is too low, while charge radii are systematically decreased by correlations for the \emarthuis{} interaction for which the mean-field predictions are too high. This feature is driven by the one-body contribution to the point-proton mean-square radius
\begin{align}
\langle R_p^2 \rangle^{\text{1b}} = \frac{1}{Z} \int d\vec{r} \, r^2 \rho_{p}(\vec{r}) \, , \label{MSpointprotonradius}
\end{align}
where the point-proton density distribution reads in the natural basis $\{\phi_{k}(\vec{r})\}$ as 
\begin{align}
\rho_{p}(\vec{r}) &= \sum_{k\in p} v^2_k \, |\phi_{k}(\vec{r})|^2 \, ,
\end{align}
with $v^2_k$ the eigenvalues of the one-body density matrix. The net effect of many-body correlations on $\langle R_p^2 \rangle^{\text{1b}}$ results from the competition between the promotion of protons from single-particle orbitals below the Fermi energy to orbitals above, the latter being typically more  extended spatially than the former, and a spatial compression of the natural orbitals that tend to reduce  $\langle R_p^2 \rangle^{\text{1b}}$. While the first effect dominates whenever mean-field radii are too small, the second effect eventually overtakes it as the mean-field baseline increases. In the future, it will be of interest to carry out a sensitivity analysis to characterize the correlation between both features and specific parameters of the nuclear Hamiltonian.

Figure~\ref{fig:tin_isotopeshift} displays isotopic shifts 
\begin{align}
\delta\langle R^{2}_{\text{ch}} \rangle^{\text{A},132} &\equiv \langle R^{2}_{\text{ch}} \rangle^{\text{A}} - \langle R^{2}_{\text{ch}} \rangle^{132} \, ,
\end{align}
computed with respect to $^{132}$Sn. Overall, the three interactions compare reasonably well with experimental data. More specifically, while \magicint{} and \emarthuis{} perform similarly ($24\%$ and $21\%$ rms error, respectively), \deltago{} does twice as good ($11\%$ rms error). 

While the sensitivity of isotopic shifts to the employed Hamiltonian is somewhat limited between $N=50$ and $N=82$, the spread of the results is strongly magnified beyond \elem{Sn}{132}, illustrating the benefit of going to such a new regime to test the performance of state-of-the-art $\chi$EFT-based inter-nucleon interactions. 

Zooming in on the \elem{Sn}{128-148} region, the right panel of Fig.~\ref{fig:tin_isotopeshift_zoom} illustrates that this divergence already impacts the capacity to reproduce the kink through \elem{Sn}{132}. Model-space uncertainties assessed by varying the HO frequency (shaded band), indicate that theoretical uncertainties are increasing with neutron-proton asymmetry  but reasonably constrained over the displayed interval. The \magicint{} and the \deltago{} Hamiltonians only account for about half of the experimental rise $\delta \langle R^{2}_{\text{ch}} \rangle^{134,132}_{\text{exp}}=0.22$\,fm$^{2}$ between \elem{Sn}{132} and \elem{Sn}{134}, which is similar to the rise $\delta \langle R^{2}_{\text{ch}} \rangle^{50,48}_{\text{exp}}$ observed through \elem{Ca}{48}. Contrarily, the sharp increase obtained for the \emarthuis{} Hamiltonian (0.23\,fm$^{2}$) reproduces accurately the experimental value. It must however be noted that the large experimental rise beyond \elem{Ca}{48} is not accounted for by the \emarthuis{} Hamiltonian and cannot actually be reproduced even when exploiting the full freedom offered by the three-nucleon interaction~\cite{Franzke:2025pvo}. 

\begin{figure*}[t!]
    \centering
    \includegraphics[height=5.65cm]{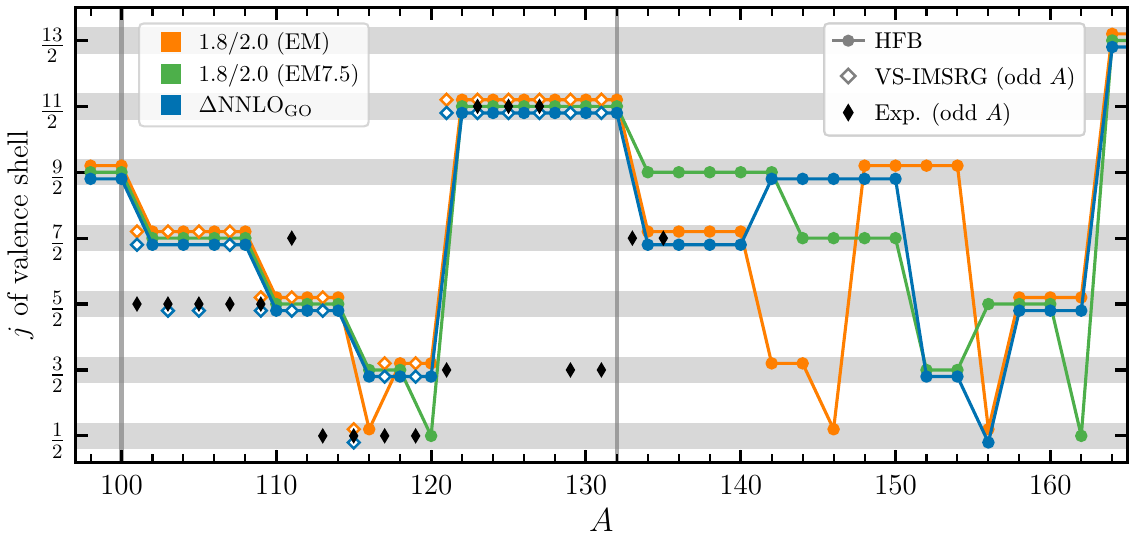}
    \hspace{1mm}
    \includegraphics[height=5.75cm]{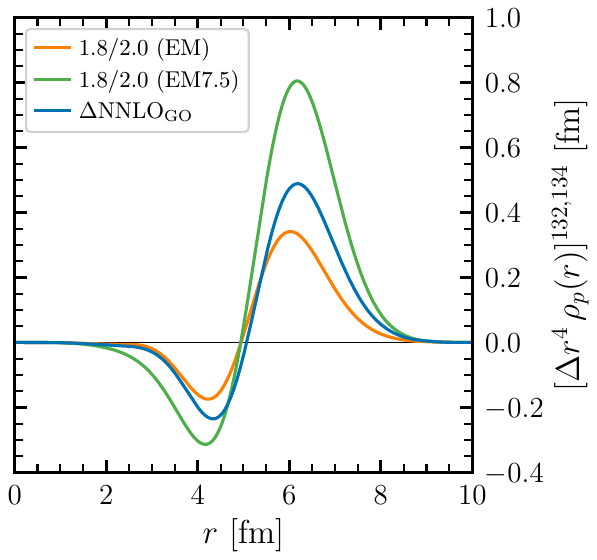}
    \caption{Left panel: angular momentum of the neutron HFB valence (canonical) shell computed for the \magicint{}, \deltago{} and \emarthuis{} interactions in even-even \elem{Sn}{98-164} isotopes (full colored circles). Ground-state angular momentum of odd-even \elem{Sn}{101-131} isotopes obtained from VS-IMSRG(2) calculations based on the \magicint{} and \deltago{} interactions (empty colored diamonds). Experimental ground-state angular momentum in odd-even \elem{Sn}{101-133} isotopes (full black diamonds). Right panel: change of the radial integrand between \elem{Sn}{132} and \elem{Sn}{134}  in the computation of the one-body  point-proton mean-square radius (Eq.~\ref{MSpointprotonradius}).}
    \label{fig:spin}
\end{figure*}

\section*{Kinks at $N=50, 82$ and $92$}

Interestingly, an inverted kink is predicted at \elem{Sn}{142} for the \emarthuis{} interaction. It is important to assess whether this observation constitutes a genuine prediction and whether it correlates with the successful reproduction of the charge-radius kink at \elem{Sn}{132}. Figure~\ref{fig:shellclosure} addresses these points by showing the differential charge radii (top) and two-neutron separation energies (bottom) around \elem{Sn}{100} (left), \elem{Sn}{132} (middle) and \elem{Sn}{142} (right). 

To make sure that the detailed patterns discussed below are robust, Fig.~\ref{fig:shellclosure} further displays model-space uncertainties. First, data points obtained by increasing $\Xmax{e}$ from $12$ to $14$ (at $\hbar\omega=12$) are included. Second, the spread of values obtained when varying $\hbar\omega$ over the interval $[10,14]$ (at $\Xmax{e} =12$) are shown as a band. While associated uncertainties of the results are negligible around \elem{Sn}{100} and \elem{Sn}{132}, they are significant around \elem{Sn}{142}. Still, the remarks and conclusions provided below are valid even when taking these non-negligible uncertainties into account.

One first observes that a large rise of the isotopic shift ($\sim0.2$\,fm$^2$) is consistently predicted between \elem{Sn}{100} and \elem{Sn}{102} for the three employed Hamiltonians. Correspondingly, the large drop of the two-neutron separation energy ($S_{2n}$) through  \elem{Sn}{100} is consistently predicted, in agreement in \elem{Sn}{102} with recent NLEFT calculations~\cite{hild25NLEFTSn}.

Moving to $N=82$, the kink in the mean-square charge radius is, as discussed above, significantly larger for the \emarthuis{} interaction than for the other two and in excellent agreement with experimental data. As for $S_{2n}$, the drop through $N=82$ is similarly exaggerated\footnote{Although not shown here, many-body correlations on top of HFB are necessary to bind Tin isotopes with respect to two-neutrons emission beyond \elem{Sn}{132} for the \magicint{} and \deltago{} Hamiltonians, thus postponing the dripline by more than 20 neutrons. } by the three Hamiltonians such that the different kinks of the isotopic shift cannot be correlated to the size of this drop. Incidentally, it will be interesting to see if the inclusion of triples excitations can attenuate the exaggerated $N=82$ magicity~\cite{vernik26}.

Finally focusing on \elem{Sn}{142}, a marked inverted kink is indeed visible for the \emarthuis{} Hamiltonian whereas the differential mean-square radius is gently varying for the other two Hamiltonians. As visible from the bottom right panel of Fig.~\ref{fig:shellclosure}, the inverted kink is correlated with a  large increase of  the $S_{2n}$ that is indeed absent for the \magicint{} and \deltago{} Hamiltonians. This dubious behavior of the $S_{2n}$ puts in question the inverted kink predicted at \elem{Sn}{142} by the \emarthuis{} Hamiltonian. 

In order to further clarify the behavior of the isotopic shifts discussed above, Fig.~\ref{fig:spin} compares the total angular momentum of the neutron valence shell extracted from the HFB calculations of even-even isotopes\footnote{The valence shell is identified via a naive filling of canonical single-particle states diagonalizing the HFB one-body density matrix and ordered according to their expectation value of the one-body Hartree-Fock field~\cite{Duguet:2020hdm}. The identification of the valence shell was found to be robust with respect to variations of $\hbar \omega \in [10,14]$.} to the ground-state angular momentum of odd-even isotopes obtained from either VS-IMSRG(2) calculations or experimental data. One first observes that ground-state angular momenta obtained from VS-IMSRG(2) calculations keep a strong memory of the underlying mean-field valence shell; the only two counter examples being for \elem{Sn}{103,105} computed with the \deltago{} Hamiltonian. While this leads to the correct value in a few isotopes (\elem{Sn}{109} and \elem{Sn}{123-127}), VS-IMSRG(2) spins are mostly at variance with experimental data. Second, while the HFB valence shell ordering is insensitive to the employed Hamiltonian between $N=50$ and $N=82$\footnote{The only exception is the inversion of the $3s_{1/2}$ and $2d_{3/2}$ sub-shells obtained when using the \emarthuis{} Hamiltonian between $N=64$ and $N=70$.}, it becomes much more versatile beyond \elem{Sn}{132}. This demonstrates how accessing heavy neutron-rich isotopes offers stringent tests to state-of-the-art $\chi$EFT Hamiltonians. Indeed, the predicted single-particle shell structure strongly impacts the evolution of mean-square radii along the isotopic chain and more specifically the three kinks discussed above
\begin{itemize}
\item {\bf \elem{Sn}{100}:} the three Hamiltonians produce the $1g_{9/2}$ shell as the neutron valence shell before $N=50$ and the significantly less bound $1g_{7/2}$ shell after the $N=50$ major shell closure. This leads to the robust prediction of the charge radius kink observed in the upper left panel of Fig.~\ref{fig:shellclosure}. However, the experimental spin in \elem{Sn}{101} is $J=5/2$ and not $J=7/2$. It is not clear whether missing many-body correlations or the Hamiltonian is responsible for this mismatch that could eventually modify the size of the predicted kink. 
\item {\bf \elem{Sn}{132}:} in agreement with the experimental spin in \elem{Sn}{133},  neutrons fill the $2f_{7/2}$ shell beyond \elem{Sn}{132} for the \magicint{} and \deltago{} Hamiltonians. Contrarily, neutrons occupy the  $1h_{9/2}$ shell for the \emarthuis{} Hamiltonian. As visible from the right panel of Fig.~\ref{fig:spin},  neutrons occupying this spatially more extended orbital pull out protons further, thus leading to the larger rise of the mean-square charge radius for the \emarthuis{} Hamiltonian in agreement with experimental data.
\item {\bf \elem{Sn}{142}:} using the \emarthuis{} Hamiltonian, the neutron valence shell beyond $N=92$ is made of the spatially less extended $2f_{7/2}$ orbitals. This produces the inverted kink of the mean square charge radius and the dubious rise of the $S_{2n}$ visible in the right panels of Fig.~\ref{fig:shellclosure}. Contrarily, no such abrupt change is observed for the  \magicint{} and \deltago{} Hamiltonians for which the less spatially extended $2f_{7/2}$ shell is filled right after $N=82$. As seen from Fig.~\ref{fig:tin_isotopeshift_zoom}, the rate at which the mean square radius increases over the interval $N \in [82,100]$ differs for both Hamiltonians. This reflects both the different binding energy of the $2f_{7/2}$ shell and the different shell ordering beyond $N=90$.
\end{itemize}
Eventually, the marked rise of the mean-square charge radius predicted by the   \emarthuis{} Hamiltonian between \elem{Sn}{132} and \elem{Sn}{134}  is fully correlated with the dubious inverted kink in \elem{Sn}{142} through the  inappropriate positioning of the  $1h_{9/2}$ shell right after $N=82$, \ie{} the reproduction of the experimental kink of the charge radius at \elem{Sn}{132} by the  \emarthuis{} Hamiltonian is obtained for the wrong reason. 

Eventually, the consistent prediction of binding energies and charge radii beyond \elem{Sn}{132} is left as a major challenge for \ai{} nuclear structure calculations and $\chi$EFT Hamiltonians.

\section*{Linear and parabolic components of $\delta \langle R^2_\text{ch} \rangle$}

\begin{figure}[t!]
    \centering
    \includegraphics[width=0.98\columnwidth]{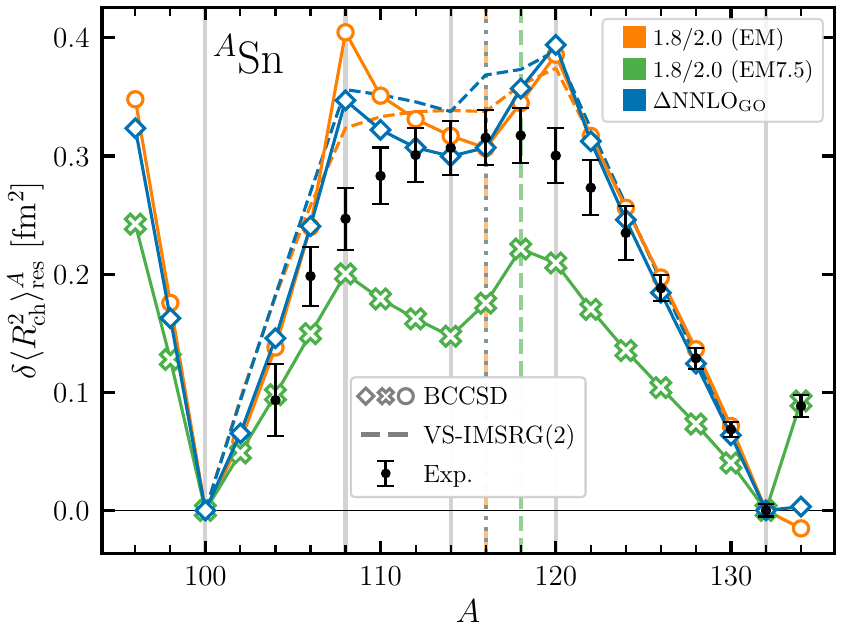}
    \caption{Experimental and theoretical (quasi) parabolic component of mean-square charge radii $\delta \langle R^2_\text{ch} \rangle^{\text{A}}_{\text{res}}$ (see text for details) between \elem{Sn}{100} and \elem{Sn}{132}. Vertical bars indicate sub-shell closures predicted at the HFB level: grey full lines specify sub-shell closures commun to the three Hamiltonians whereas colored dashed and dashed-dotted lines indicate those that are interaction specific in connection with the inversion of  the $3s_{1/2}$ and $2d_{3/2}$ shells for the \emarthuis{} Hamiltonian; see Fig.~\ref{fig:spin} for more details. }
    \label{fig:dRres}
\end{figure}

The behavior of the isotopic shifts seen in Fig.~\ref{fig:tin_isotopeshift} between \elem{Sn}{100} and \elem{Sn}{132} can be well decomposed into a linear component plus a quasi-parabolic one. For both experiment and theory, this quasi-parabolic component
\begin{align}
\delta \langle R^2_\text{ch} \rangle^{\text{A}}_{\text{res}} \equiv \delta \langle R^2_\text{ch} \rangle^{\text{A},132} - a(A-132) \, ,
\end{align}
where
\begin{align}
a \equiv - \frac{\delta \langle R^2_\text{ch} \rangle^{100,132}}{32} \, ,
\end{align}
is displayed in Fig.~\ref{fig:dRres}. Since experimental mean-square radii are not known below \elem{Sn}{104}, the value at \elem{Sn}{100} necessary to compute the slope $a$ is obtained by fitting experimentally known isotopic shifts by the sum of a linear and a quadratic term\footnote{Notice that experimental mean-square radii are actually best reproduced by combining two successive parabola over the intervals $N=50-64$ and $N=64-82$~\cite{gustaf21phd}.} \cite{kart24In}. 

As can already be appreciated from  Fig.~\ref{fig:tin_isotopeshift}, the linear slope from experimental data ($a=0.068$\,fm$^2$) is well accounted for by the \deltago{} and \emarthuis{} Hamiltonians ($a=0.069$\,fm$^2$ and $0.067$\,fm$^2$, respectively) but much less by the \magicint{} Hamiltonian ($a=0.061$\,fm$^2$). These BCCSD values are  essentially identical to their VS-IMSRG(2) counterparts. 

As shown in Fig.~\ref{fig:dRres}, the quasi-parabolic component of experimental mean-square radii is overall well reproduced by the \magicint{} and \deltago{} Hamiltonians, whereas it is significantly underestimated by the \emarthuis{} Hamiltonian. Looking more closely, the underlying subshell closures leave distinct fingerprints in BCCSD results in the form of local maxima and minima that are not visible in the experimental data. Moving to VS-IMSRG(2), such local irregularities are significantly smoothed out but remain somewhat visible, pointing to yet missing, probably collective, many-body correlations.

\section*{Conclusions and outlook}

Charge radii along the Tin isotopic chain have been studied on the basis of \ai{} Bogoliubov coupled cluster calculations at the singles and doubles level. The rich features displayed by isotopic shifts along this large sequence of semi-magic isotopes, \ie{} the linear and parabolic components between the $N=50$ and $N=82$ magic numbers and the kink through \elem{Sn}{132}, have been shown to provide stringent tests for state-of-the-art $\chi$EFT inter-nucleon interactions. Indeed, none of the three fine-tuned Hamiltonians employed in the present study can capture all key characteristics. While a significant kink is robustly predicted through \elem{Sn}{100}, the experimentally known kink through \elem{Sn}{132} could only be reproduced at the price of a dubious predicted shell structure beyond the $N=82$ shell closure. Eventually, the increased sensitivity of the results to the employed Hamiltonian beyond \elem{Sn}{132} offers a potentially fruitful playground to pin down critical attributes of $\chi$EFT inter-nucleon interactions.

In this context, it would be highly beneficial to extend the experimental measurement of isotopic shifts both towards \elem{Sn}{100} and beyond \elem{Sn}{134}. Moreover, it is of prime interest to perform a thorough sensitivity analysis of the presently discussed results with respect to the low-energy constants of the employed  $\chi$EFT Hamiltonian in the spirit of, \eg{}, Ref.~\cite{Franzke:2025pvo}. Finally, it will be necessary to perform high-accuracy \ai{} calculations of mean-square radii in heavy open-shell nuclei in the forthcoming future. This is to be done by both adding triples correction to BCCSD computations~\cite{vernik26}, which is expected to be on the percent level~\cite{Bon26privcom}, and the two-body charge density correction that has been shown to contribute to about $0.04$\,fm in $p$-shell nuclei~\cite{Sun:2026eep}. One must also contemplate the need to include truly collective collective correlations that are typically hard to grasp via single-reference expansion methods based on state-of-the-art truncations or  valence-space methods based on standard closed cores.

\section*{Acknowledgements}

The authors thank P.~Arthuis, F.~Gustafsson, T.~Miyagi, G.~Neyens, T.~Papenbrock and A.~Schwenk for useful discussions, as well as T.~Miyagi for providing VS-IMSRG(2) results. This work was supported in part by the LOEWE Top Professorship LOEWE/4a/519/05.00.002(0014)98 by the State of Hesse, by the European Research Council (ERC) under the European Union's Horizon Europe research and innovation programme (Grant Agreement No.~101162059), and by Research Foundation Flanders (FWO, Belgium, grant 11G5123N). 
The present research benefited from computational resources made available on Lucia, the Tier-1 supercomputer of the Walloon Region, infrastructure funded by the Walloon Region under the grant agreement n°1910247.
The authors gratefully acknowledge the Gauss Centre for Supercomputing e.V. (www.gauss-centre.eu) for funding this project by providing computing time through the John von Neumann Institute for Computing (NIC) on the GCS Supercomputer JUWELS at J\"ulich Supercomputing Centre (JSC).

\bibliographystyle{apsrev4-2}
\bibliography{strongint}

\end{document}